# A fast-rotating pear-shaped nucleus

A. Karmakar, P. Datta, Soumik Bhattacharya, Shabir Dar, S. Bhattacharya, G. Mukherjee, H. Pai, S. Basu, S. Nandi, S. S. Nayak, S. Das, R. Raut, S.S. Ghugre, R. Banik, Sajad Ali, W. Shaikh, G. Gangopadhyay, and S. Chattopadhyay

## Abstract

The spectroscopic studies have identified few Actinide and Lanthanide nuclei of the periodic table, which can assume the pear shape. The low frequency collective rotation of these nuclei has been established by determining the band structure of the excited levels and their gamma decay rates. In this article, we report the rotation of a pear-shaped nucleus in $^{100}$Ru, which rotates nearly three times faster than the previously known cases. The three novel consequences of this fast rotation are: the realization of a pear-shape in an excited state, its moment of inertia becoming a constant of motion and its shape evolution. These inferences have been arrived at by comparing the characteristics of the rotational band of $^{100}$Ru with three of the best-known examples of a rotating pear-shaped nucleus and through the theoretical interpretation of the data. The observation of a pear-shaped nucleus in the lighter mass region opens up the possibility of a systematic study of the effects of fast rotation on this shape.

## Introduction

The nuclear rotation was first described by Bohr and Mottelson[1] by considering the nucleus as a quantum droplet which can assume non-spherical shapes due to the shell structure. The non-spherical shape changes its orientation under rotation, which is the essential criterion for quantum rotation. It is now established that the nuclei with the proton and/or the neutron numbers well separated from the magic numbers exhibit quadrupole deformation (characterized by the deformation parameter $\beta_2$), which can be axially symmetric prolate (lemon-shaped) and oblate (orange-shaped) shapes or asymmetric triaxial (shape of a mango seed) shape. A pear shape can be realized by superimposing octupole deformation (characterized by $\beta_3$) on the prolate shape. This shape has been reported only in few Actinide and the Lanthanide nuclei, where it originates due to the long-range octupole-octupole correlations among the nucleon orbitals close to the Fermi surface, whose angular momentum (J) differ by 3 ℏ. The rotation of a pear-shaped even-even (proton-neutron) nucleus exhibits a unique sequence of interspaced states, whose angular momentum (analogously called 'spin' in the text) and parity are $I^\pi = 0^+, 1^-, 2^+, 3^-, 4^+, 5^-, 6^+, 7^-$ ... and they are connected by relatively fast electric dipole (E1) transitions ($I^\pm \rightarrow (I-1)^\mp$). At the same time, the states of the same parity are connected by electric quadrupole (E2) transitions ($I^\pm \rightarrow (I-2)^\pm$) thereby forming two parity partner bands. Such a sequence was first reported in $^{218}$Ra[2] and since then, the octupole deformation has been reported in a number of even-even isotopes of Ra – Th[3-5] and Sm – Ba nuclei[6-9]. The origin of the parity partner bands can be understood by plotting the nuclear potential energy of this reflection-asymmetric shape as a function of octupole deformation[3, 10-12]. This is shown in the inset (a) of Fig 1. The potential energy has two degenerate minima at $\pm \beta_3^{min}$ separated by a finite barrier at $\beta_3 = 0$. The two parity partner bands are associated with the rotations of the two mirror shapes in these two minima. However, if the barrier height is finite, tunnelling is possible between the two minima. This leads to the observed parity splitting in the laboratory frame. The other fingerprint of the octupole deformation is the relative enhancement in the E1 transition rates. This was first considered by Bohr and Mottelson[1] in the framework of the liquid drop model. In this macroscopic picture, the concentration of protons is more in the region of higher curvature (lighting rod effect), which is the narrower end of the pear. This leads to a separation between the centre of charge and the centre of mass thereby imparting an intrinsic electric dipole moment, which leads to an interleaved sequence of relatively fast E1 transitions between the partner bands. The review articles[10-13] give a comprehensive account of the experimental and theoretical studies on the reflection asymmetric nuclei.

There exists another mass region, namely A ~ 96 (Z ~ 40 and N ~ 56), where the neutron and the proton orbitals satisfy the $\Delta J = 3\hbar$ criterion like in the case of Ra – Th and Sm – Ba nuclei. In this mass region, it is possible to observe the fast rotation of a pear-shaped nucleus as these nuclei are substantially lighter. This is an interesting prospect as the phenomena like nuclear shape evolution with possible occurrence of triaxial shapes[14] and rigid nuclear rotation[15], have been observed in the fast-rotating nuclei with quadrupole deformation. However, the consequences of fast rotation on pear-shaped nuclei remained unexplored as no such light nucleus could be identified in the last three decades of experimental investigations and a recent systematic theoretical study indicated that the nuclei of A ~96 region are unlikely of possess ground state octupole deformation[16]. In this mass region the two N = 56 isotones namely, $^{96}$Zr [17] (Z = 40) and $^{98}$Mo [18] (Z = 42), exhibit octupole instability but the rotational band structures are absent. On the other hand, $^{100}$Ru (Z = 44) has two interspaced rotational bands of opposite parity[17] beyond I = 11 $\hbar$ thereby indicating the possibility of the presence of an excited octupole band but the interleaved E1 transitions were not observed. Since $^{100}$Ru presents this unique possibility, we decided to re-populate its excited levels through the fusion-evaporation reaction involving the 50 MeV alpha beam on the $^{100}$Mo target. This reaction pre-dominantly populates the levels around I ~ 12-14 $\hbar$, where we planned to search for these elusive E1 transitions by collecting a large data sample.

# Experimental data and discussion

The relevant excited levels of $^{100}$Ru labelled by the energy (E) and the spin-parity ($I^\pi$) (commonly known as 'partial level scheme') are shown in Fig. 1, where the thicknesses of the arrows are proportional to the relative intensities of the deexciting gamma rays. The newly placed transitions from the present data are marked in red. These placements have been confirmed through the observed coincidences and anti-coincidences with the known gamma rays[19-20]. In this figure, four rotational band structures can be observed. The excitation energy (E) of the levels of a rotational band has the characteristic spin (I) dependence given by $E(I) \propto I(I+1)$, where the proportionality constant is the inverse of the moment of inertia (MOI). At low spin, the MOI of an even-even nucleus (like $^{100}$Ru) is considerably smaller than the classical rotor value due to the long-range pairing correlations among the nucleons. These correlations originate due to the scattering of a pair of nucleons between the orbitals near the Fermi surface. As the nucleus starts to rotate faster, the motions of the nucleons in the intrinsic frame are strongly influenced by the Coriolis force, which breaks the nucleon pairs in the time reversed orbitals and starts to align them along the rotation axis (commonly known as 'rotational alignment'). Thus, in general, the angular momentum of a nucleus gets generated by the contributions from both the collective rotation and the orbital angular momenta of the aligning nucleons and the MOI increases with increasing angular momentum.

Band1 of Fig. 1 (levels marked in green) is the ground state band of $^{100}$Ru, where all the nucleons occupy the orbitals pairwise and thus, the pairing correlations are strong in this band and the MOI is low. Band4 (levels marked in magenta) on the other hand, is a negative parity odd spin band. This rotational band in $^{100}$Ru can arise if a nucleon pair is broken and one of them is excited to a higher orbital such that the two nucleons occupy opposite parity orbitals. For the deformed nuclei of this mass region, the neutron $h_{11/2}$ (as defined in the Nuclear Shell model) is the only available negative parity orbital near the Fermi surface. Thus, this rotational band may arise due to a broken pair of neutrons. In this case, the neutron pairing correlations become small due to Pauli blocking in the neutron sector and as observed from Fig. 1, the MOI of Band4 is higher compared to Band1.

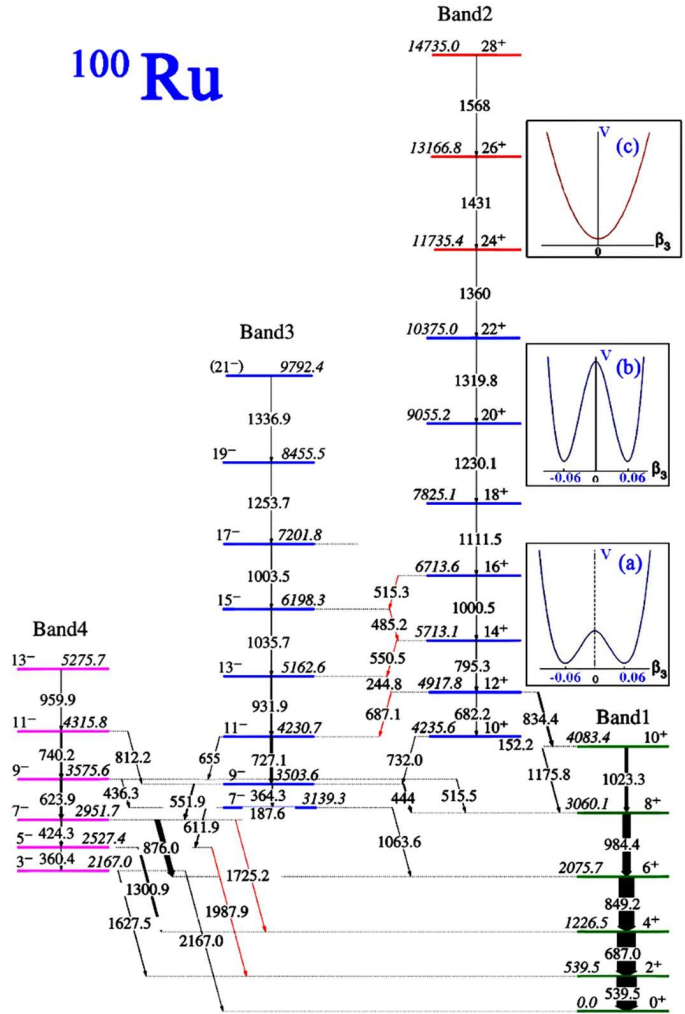

**Fig. 1** The partial level scheme of $^{100}$Ru established from the present work and the evolution of the nuclear potential energy with increasing angular momentum. The seven gamma transitions marked in red, were observed during the present work. All other transitions and the spin-parity of the excited levels have been previously reported in ref. 19 and 20. The widths of the arrows are proportional to the relative intensities. The level and transition energies are expressed in keV and the uncertainties in the transition energies are ± 0.2 keV. A schematic representation of the evolution of the nuclear potential energy of $^{100}$Ru with increasing angular momentum is shown in the three insets. The plotted values correspond to $\beta_2 = 0.18$ and $\beta_3 = 0.06$. The three situations are as follows: (a) the region of stable octupole deformation ($11\hbar \leq I \leq 16\hbar$), where the barrier height at $\beta_3 = 0$ is small and the tunnelling between the two minima is present; (b) for $16\hbar < I \leq 22\hbar$, the barrier height becomes large due to substantial loss in the pairing correlations and the tunnelling stops; (c) at high angular momentum ($I > 22\hbar$), the nucleus changes to a reflection symmetric shape, which has been depicted by the potential energy minimum at $\beta_3 = 0$.

The most significant feature of the present level scheme (Fig. 1) is the presence of the two interspaced opposite parity bands (Band2 and Band3) interconnected through five E1 transitions, thereby suggesting them as the parity partners of an octupole band (levels marked in blue). The presence of the 732 keV transition from Band2 ($10^+ \to 9^-$) indicates the presence of the octupole correlations since this transition would have been absent for the case of octupole vibration (dynamic octupole deformation)[21]. These correlations also manifest in the enhanced E1 transition rate of the 444 keV transition from the $9^-$ level of Band3 to Band1. This inter-band E1 rate normalized by the intra-band B(E2) rate is $0.10(2) \times 10^{-6}$ fm$^{-2}$, whereas the corresponding value for the 516 keV transition is $0.04(1) \times 10^{-6}$ fm$^{-2}$ for the $9^-$ level of Band4, which originates due to the single particle excitation. Thus, it may be concluded that the lower spin states of the parity doublet bands exhibit the presence of the octupole correlations, which lead to the stable octupole deformation beyond I = 11 $\hbar$.

In order to establish the rotation of the pear-shaped $^{100}$Ru, the ratio of the dipole ($D_0$) and the quadrupole transition moments ($Q_0$) has been extracted from the measured intensity ratio of the feed-out E1 and E2 transitions for the different levels of the octupole band and plotted in Fig. 2. If we assume $Q_0 \sim 200$ e fm$^2$ for $^{100}$Ru following the systematics of the B(E2; $2^+ \to 0^+$) transition rate[22], the intrinsic dipole moment turns out to be $\sim 0.04$ e fm by taking an average of the $D_0/Q_0$. This $D_0$ value corresponds to an average B(E1) retardation of $\sim 10^4$ for the levels of the two parity partners. This rate is an order of magnitude higher than the commonly observed B(E1) rates of this mass region and those estimated for the low spin levels of Band4. A similar order of magnitude increase in the B(E1) rates has also been reported in the light Ra and Th isotopes. Thus, the presence of the interspaced opposite parity levels along with the relatively fast interleaved E1 transitions, confirms that the levels beyond I = 11$\hbar$ form a stable octupole band in $^{100}$Ru and thereby, establishes it as the lightest pear-shaped nucleus of the periodic table. In Fig. 2, the $D_0/Q_0$ values for $^{226}$Th[23] have been plotted, which is one of the best examples of a rotating pear-shaped nucleus with $\beta_3 \sim 0.1$ [5]. It is evident from the figure that the average $D_0/Q_0$ value for $^{226}$Th is about a factor of two higher than the value observed for $^{100}$Ru. This observation indicates that the octupole deformation parameter ($\beta_3$) of $^{100}$Ru is about 0.06 as estimated by using the A$^{1/3}$ mass-scaling[24]. The graphical representation of the shape of $^{100}$Ru is shown in Fig. 3 along with $^{226}$Th, whose shape parameters have been reported in ref. 25.

It is observed from Fig. 1 that the smooth band structure (E(I) $\propto$ I(I+1)) of Band1 exhibits a sharp discontinuity beyond I = 10 $\hbar$ and a new band (Band2) becomes favoured in energy from I = 12 $\hbar$. This phenomenon[26] (commonly known as 'back-bending') happens, when the Coriolis force decouples a pair of nucleons and aligns their angular momentum vectors along the rotational axis. In this case, their angular momenta add to form a high spin state at an energy lower than needed to generate the same spin through the collective rotation. For the prolate nuclei of this mass region, the observed back-bending around I = 10 ℏ corresponds to the alignment of two $h_{11/2}$ neutrons[14]. This systematic indicates that the octupole band in $^{100}$Ru originates from an intrinsic configuration with two rotationally aligned neutrons, which is pictorially shown in Fig. 3(a). It is interesting to note that all the previously known octupole bands in even-even nuclei are based on the ground state configuration, where all the nucleons are paired and the higher angular momenta are predominantly generated through the collective rotation of the pear-shaped nucleus. This situation is depicted in Fig. 3(b).

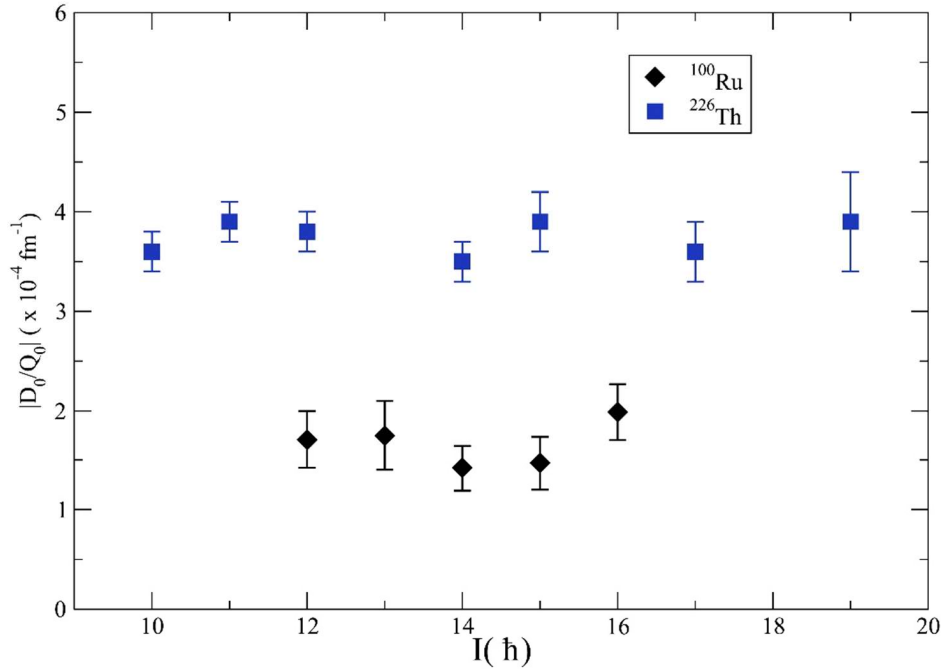

**Fig. 2** The ratios of the transition dipole and quadrupole moments extracted from the experimental branching ratios have been plotted as a function of spin, I, for $^{100}$Ru and $^{226}$Th[23].

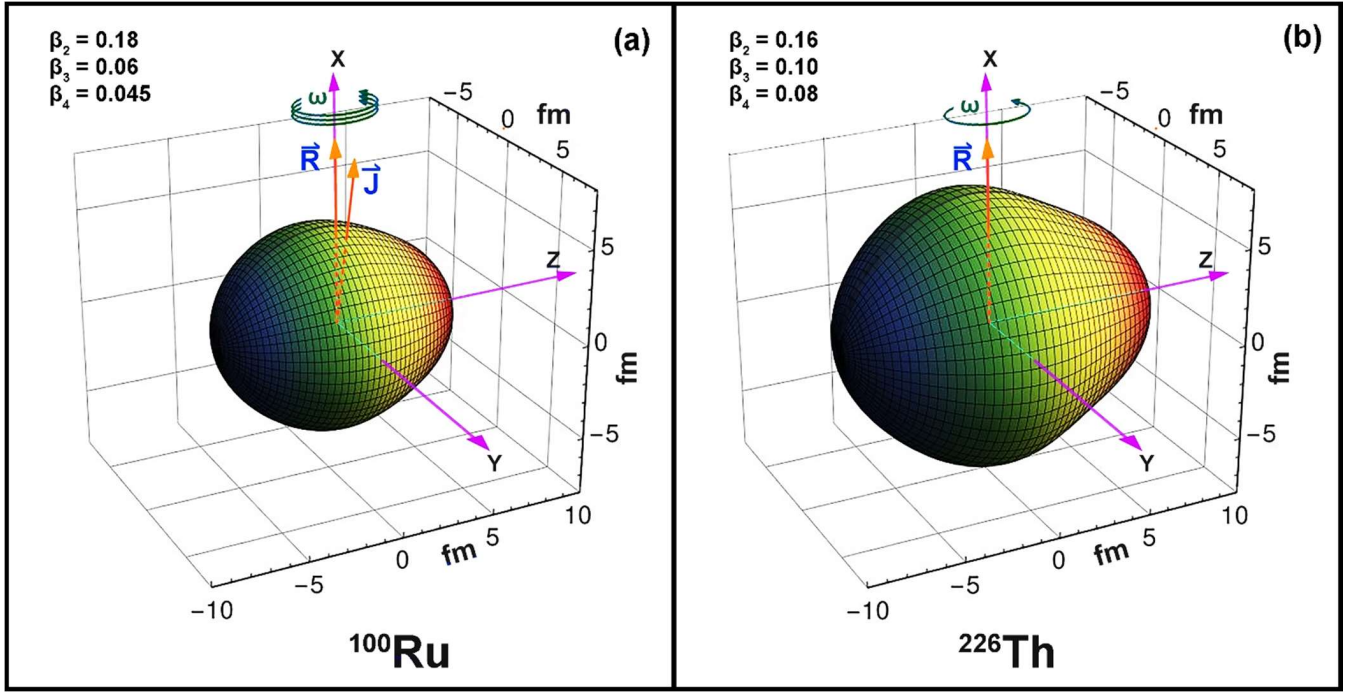

**Fig. 3** Panels (a) and (b) represents the shapes of $^{100}$Ru and $^{226}$Th, respectively. The shape parameters for $^{226}$Th has been taken from ref. 25. It may be noted that $^{226}$Ra also has similar shape. The estimate for the $\beta_3$ value for $^{100}$Ru has been obtained from the observed B(E1)/B(E2) rates (as explained in the text), while the values for $\beta_2$ and $\beta_4$ have been obtained from the self-consistent HFODD calculations. The smaller $^{100}$Ru rotates nearly three times faster than $^{226}$Th in order to generate the same angular momentum. The pear shape of $^{100}$Ru is associated with two rotationally alignment neutrons in the intrinsic frame and their resultant angular momentum vector is presented by **J**, while **R** is the rotational angular momentum vector. For the two aligned neutrons |**J**| ~ 10 ℏ and the rotation of the pear-shaped $^{100}$Ru is observed only beyond this spin value. For the highest observed rotational level of $^{100}$Ru, |**R**| ~ 10 ℏ. In case of $^{226}$Th (and for all the previously known even-even pear-shaped nuclei), the angular momentum of ~ 20 ℏ is predominantly generated by collective rotation alone. This situation has been depicted in panel (b).

In order to further explore the properties of this unique octupole band, the MOI ($J^{(1)}$) of the two parity partners of $^{226}$Ra, $^{226}$Th, $^{144}$Ba and $^{100}$Ru have been determined from the usual expression of quantum rotation and plotted on the left panel of Fig. 4. The rotational frequency (ω) of an even-even nucleus, is defined as half of the transition energy between two levels [(I+1) → (I-1)] and is identified with the mean value (I). On the right panel, the parity splitting indices, S(I$^+$) and S(I$^-$), for the positive and negative parity bands have been plotted, where S(I) is defined as the difference of the energy difference of the I, (I - 1) and (I – 2) levels[14]. The first distinguishable feature observed from the plots is that $^{100}$Ru rotates nearly three times faster than the nuclei of the A ~ 222 region. The consequences of this high frequency rotation of a pear-shaped nucleus have been established through the comparison plots of Fig. 4.

Fig. 4(a) shows that at lower frequencies, the negative parity band of $^{226}$Ra has a significantly higher moment of inertia due to the mixing with the bands with one broken pair of nucleons. However, with increasing frequency, the MOI of the positive parity band starts to increase sharply due to the Coriolis force and at a higher frequency (ħω ~ 0.18 MeV), both of the members of the parity doublet have similar MOI (shown in red and black in Fig. 4(a)), leading to the onset of strong octupole correlations. This drives the nucleus to a stable octupole shape. This transition has been quantified by the critical angular momentum (I$_c$) for which S(I$^+$) = S(I$^-$) = 0 [27]. For $^{226}$Ra, I$_c$ = 13 ħ (from Fig. 4(b)), which corresponds to ħω = 0.18 MeV. With the onset of stable octupole deformation, the negative parity partner becomes favoured in energy (Fig. 4 (b)) and the parity splitting induces a small splitting in the MOI values of the two partners indicating the presence of the tunnelling between the two minima through the potential barrier at β$_3$ = 0. With the increasing rotational frequency, the parity splitting vanishes beyond I = 23 ħ as the pairing correlations become weaker and the barrier height increases[28]. Thus, at low rotational frequencies $^{226}$Ra exhibits octupole instability due to the presence of the large pairing correlations, which is also reflected in the substantially different MOI values for the two parity partners; in the intermediate rotational frequencies the pairing correlations decreases and the octupole deformation stabilizes but the parity splitting persists; while at higher spins, the two asymmetric shapes rotate independently separated by the large potential barrier at β$_3$ = 0 and the parity splitting vanishes (represented by the inset (b) of Fig. 1). This may be understood from the fact that the pairing correlations respect the π-rotational symmetry of an axially symmetric even-even nucleus when rotated perpendicular to the symmetry axis, while the reflection-asymmetric pear shape breaks this symmetry.

In the case of $^{226}$Th, the region of instability persists longer where the values of the MOI of the partners are distinct and S(I$^-$) is unfavoured in energy (Fig. 4 (c) and (d)). Beyond I$_c$ = 15 ħ (ħω = 0.20 MeV), the S(I) values remain zero, which is indicative of a barrier height large enough to stop the tunnelling. It is interesting to note that the intermediate domain of parity splitting is absent in this case.

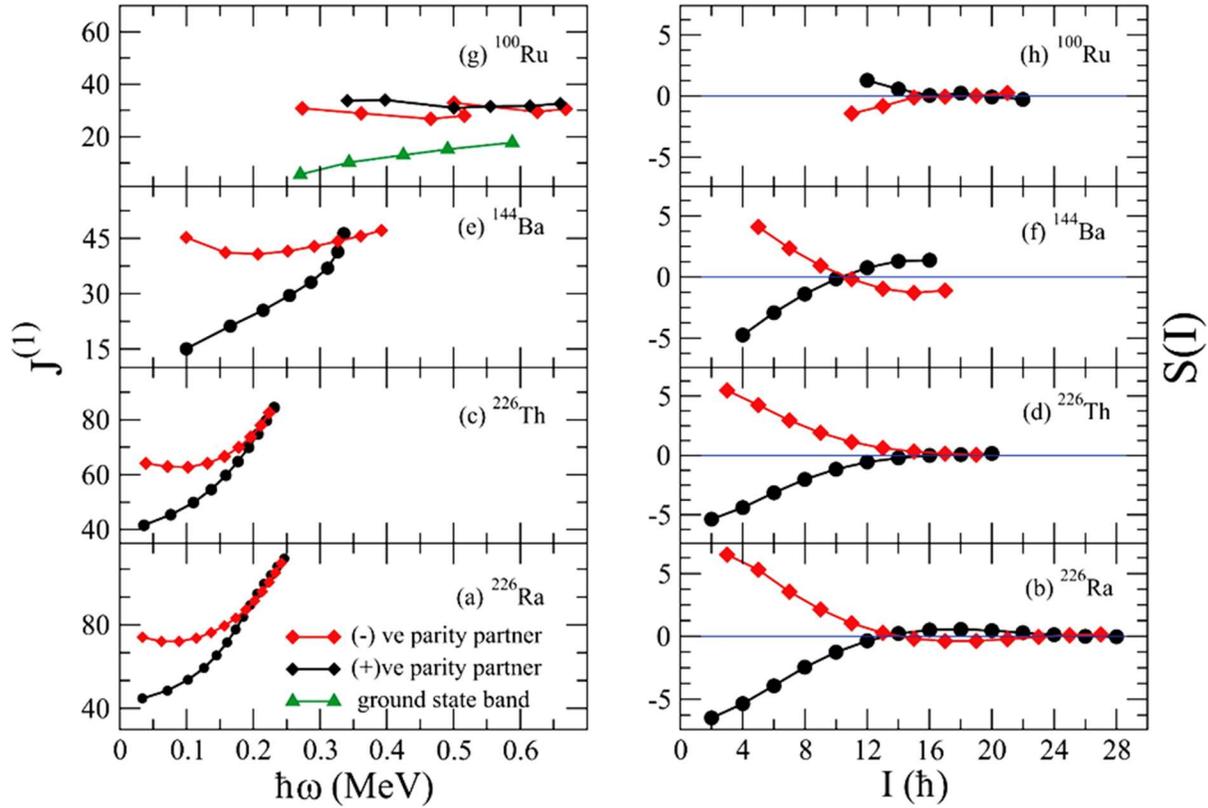

**Fig. 4** The moments of inertia ($J^{(1)}$) and the S(I) variable for the positive (black dots) and the negative parity partner (red dots) bands are plotted on the left and the right panels as a function of rotational frequency (ω) and angular momentum, respectively, for $^{226}$Ra, $^{226}$Th, $^{144}$Ba and $^{100}$Ru. The $J^{(1)}$ values for the ground state band of $^{100}$Ru has been plotted in green. The connecting lines have been drawn purely for eye guidance as the values of the variables are discrete.

For $^{144}$Ba, the HFB calculations[6] predict a phase transition from octupole vibration to octupole deformation at I = 10ℏ. This onset of the octupole deformation is clearly visible at $I_c$ = 10 ℏ in the S(I) plot of Fig. 4(f). Beyond this point (ℏω = 0.29 MeV), there is a large parity splitting, which induces a large separation of the MOI values of the two partners as seen in Fig. 4(e). This trend continues till the highest observed spin of I = 17 ℏ.

The MOI and S(I) plots for $^{100}$Ru are distinct from all the other cases as seen in Fig. 4(g) and (h). The larger rotational frequency facilitates the rotational alignment of two neutrons. The presence of the two unpaired neutrons in the immediate vicinity of the Fermi surface leads to a substantial loss in the pairing correlations in the neutron sector due to Pauli blocking. This is reflected in the large values of the moments of inertia of two partner bands compared to the value for the ground state band (shown in green in Fig. 4(g)). The S(I) plot (Fig. 4(h)) shows a stable octupole deformation for $^{100}$Ru and the region of octupole instability observed in $^{226}$Ra and $^{226}$Th, is absent. This difference is due to the presence of a strong pairing correlations in the positive parity partner bands of $^{226}$Ra and $^{226}$Th, which is substantially quenched in the rotationally-coupled octupole band of $^{100}$Ru. The small splitting in the MOI values in $^{100}$Ru is linked to the parity splitting observed till I = 15 ℏ, as observed in Fig. 4(h) (similar to $^{226}$Ra). Beyond I = 15 ℏ, (ℏω ≥ 0.5 MeV), the splitting of the parity and MOI values vanish as at the higher rotational frequencies the pairing correlations of the protons are also expected to get quenched. Thus, in the high frequency domain, the two mirror shapes of $^{100}$Ru rotate independently, as observed in $^{226}$Ra and $^{226}$Th. The evolution of the potential barrier at $\beta_3 = 0$ in these two-spin domains has been depicted in the insets (a) and (b) of Fig. 1. It is interesting to note from Fig. 4(g) that for ℏω ≥ 0.5 MeV, the moments of inertia of the two parity partner bands of $^{100}$Ru become independent of the rotational frequency. This implies that the high spin states (15ℏ ≤ I ≤ 22ℏ) are generated only through the collective rotation of this pear-shaped nucleus.

In order to investigate the effects of the high frequency rotation of $^{100}$Ru within a theoretical framework, we have calculated the strength of the pairing correlation as a function of the rotational frequency with and without octupole deformation. These numerical calculations were performed using HFODD code[29], which uses the deformed harmonic oscillator basis and solves the nuclear Hartree-Fock-Bogolyubov (HFB) problem for zero-range Skyrme interactions. The strengths of the proton and neutron pairing forces have been chosen so that the ground state solution for the nucleus reproduces the experimental pairing gaps, which have been obtained through the three-point formula of odd even mass difference. The octupole deformation value has been chosen to be $\beta_3 = 0.06$, which is consistent with other mass regions after the mass scaling. Apart from fixing the octupole deformation, the solutions for all the rotational frequencies have been obtained self-consistently without any symmetry restriction. We find that the quadrupole deformation in all the cases is around $\beta_2 = 0.18$. The proton and the neutron pairing correlations (represented by the pairing gap values) obtained from the present self-consistent calculations have been plotted as a function of the rotational frequency in Fig. 5. It is interesting to note from Fig. 5(a) that the pair gap energy values for the neutrons is lower at all frequencies for $\beta_3 = 0.06$ (red curve) compared to the values for $\beta_3 = 0$ and the neutron alignment takes place at a lower frequency of 0.45 for $\beta_3 = 0.06$. This crossing frequency is consistent with the observed rotational alignment frequency of the neutrons. Thus, these calculations seem to indicate that the rotationally aligned neutron configuration is favoured in energy for the pear-shaped $^{100}$Ru. On the other hand, the pair gap energy values for the protons are identical for frequencies below 0.5 MeV with and without octupole deformation. These contrasting behaviours indicate strong octupole interactions among the neutron orbitals near the Fermi surface at N = 56, which are otherwise absent in the proton sector (Z = 44). At higher frequencies, the proton pairing correlations vanish around $\hbar\omega$ = 0.55 MeV for $\beta_3$ = 0.0 (blue curve in Fig 5(b)), which corresponds to the alignment of two protons. For $\beta_3$ = 0.06 (the red curve), the correlation strength starts to decrease sharply only beyond 0.5 MeV, which is consistent with the observation of the vanishing of parity splitting at higher rotational frequencies. In addition, the nearly 75% decrease in the pairing correlations (neutron + proton) beyond $\hbar\omega$ = 0.5 MeV, as observed from Fig. 5, is the probable reason for the observation of the pure collective rotation in $^{100}$Ru in this domain.

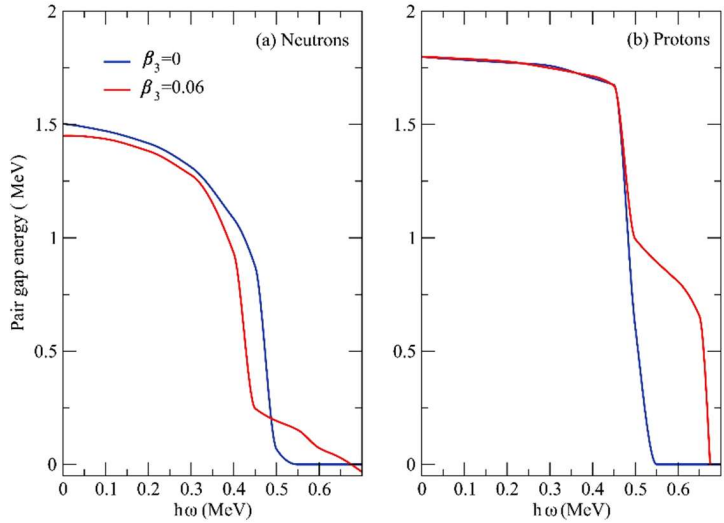

**Fig. 5** The variation of the neutron and proton pair gap energies as a function of the rotational frequency in $^{100}$Ru. The pair gap energy is the measure of the pairing correlation and has been calculated self-consistently using the HFODD code for (a) the neutrons and (b) the protons. The pair gap energy has been calculated in steps of 0.05 MeV by solving the pair-gap equation. The curves correspond to the spline fits to the calculated values. It is seen from (a) that the neutron pair gap energy for $\beta_3$ = 0.06 at $\hbar\omega$ = 0.45 is nearly four times smaller than that for $\beta_3$ = 0 thereby indicating an earlier rotational alignment of neutrons for a pear shaped $^{100}$Ru in comparison to a lemon shape.

It is observed from Fig. 5 that the pairing correlations in $^{100}$Ru vanishes completely beyond 0.65 MeV for $\beta_3$ = 0.06 and 0.55 MeV for $\beta_3$ = 0. This is because at higher frequencies the stronger Coriolis force mixes the different mixed-parity orbitals, which originate from the $h_{11/2}$ ($g_{9/2}$) and the $d_{5/2}$ ($p_{3/2}$) for the neutrons (protons), due to the presence of the octupole deformation. This contamination by the low-j orbital delays the alignment frequencies in pear-shaped nuclei [2, 11-13]. The complete self-consistent calculations for $\beta_3 = 0$, indicate that the high frequency rotation of $^{100}$Ru favours a triaxial shape after the vanishing of the pairing correlations. It is interesting to note from Fig. 1 that beyond I = 22 $\hbar$ (corresponding to $\hbar\omega$ = 0.66 MeV) the parity doublet structure has been found to be lost as Band3 terminates abruptly and only Band2 extends to higher spins. However, the band structure of Band2 exhibits a discontinuity after 22 $\hbar$ and a new band structure arises beyond this spin. This high spin band may probably be associated with the reflection-symmetric triaxial shape observed in the HFODD calculations in the absence of pairing correlations. However, it is to be noted that the present framework cannot explore the transition of $^{100}$Ru from a pear to a triaxial shape as the value of the octupole deformation has been fixed and not obtained from the self-consistent calculations. It may be noted that a similar sudden termination of an octupole band has been reported in $^{222}$Th[30], where a reflection-symmetric band with rotationally-aligned two neutrons and two protons has been predicted[31] to become yrast beyond I = 25 $\hbar$. Thus, the abrupt termination of Band3 can be associated with the shape evolution of the pear-shaped $^{100}$Ru to a reflection-symmetric shape at high rotational frequencies.

## Summary

We have observed the fast rotation of a pear-shaped nucleus and its consequences for the first time. We have also reported the first observation of an excited octupole band. The numerical calculations using HFODD code indicate that the pear shape of $^{100}$Ru is energetically favoured for the rotational alignment of a pair of neutrons in comparison to a lemon shape. At high rotational frequencies, the moment of inertia of $^{100}$Ru remains constant due to the significant decrease in the pairing correlations as inferred from the calculations. At even higher rotational frequencies, the present work suggests that $^{100}$Ru evolves from a pear to a reflection-symmetric shape, which is the plausible reason for the abrupt termination of the octupole band. The systematic study of the parity doublet bands of $^{100}$Ru, $^{144}$Ba, $^{226}$Ra and $^{226}$Th, establishes the fact that the octupole deformation stabilizes with the decrease in the pairing correlation. It will be interesting to extend this study to the other N=56 isotones, which may also be pear shaped due to the rotational alignment of a pair of neutrons. These studies may reveal more exotic features, for example, the elusive rigid rotation of a pear-shaped nucleus or the rotation of a heart-shaped nucleus.

# Acknowledgements

The authors convey sincere thanks to the operational staff of the K-130 cyclotron at VECC for providing good quality beam as well as necessary support during the pandemic period. The authors are thankful to Department of Atomic Energy and Department of Science and Technology, Government of India for providing the necessary funding for the funding for the Clover array. A.K. acknowledges the Council of Scientific Research (CSIR) (File No: 09/489(0121)/2019-EMR-I), Government of India.# Author information

*Authors and Affiliations*

*Saha Institute of Nuclear Physics, Kolkata India and Homi Bhabha National Institute, Mumbai, India.*

A. Karmakar, H. Pai* and S. Chattopadhyay[1]

* Present address: Extreme Light Infrastructure - Nuclear Physics, Horia Hulubei National Institute for R&D in Physics and Nuclear Engineering, Bucharest-Magurele, 077125, Romania.

*Ananda Mohan College, Kolkata- 700009, India.*

P. Datta

*Variable Energy Cyclotron Centre, Kolkata, India and Homi Bhabha National Institute, Mumbai, India.*

Soumik Bhattacharya**, Shabir Dar, S. Bhattacharya, G. Mukherjee, S. Basu, S. Nandi[#], S. S. Nayak, S. Das

** Present address: Department of Physics, Florida State University, Tallahassee, Florida, USA.

[#] Present address: Physics Division, Argonne National Laboratory, Lemont, Illinois 60439, USA.

*UGC-DAE CSR, Kolkata, India.*

R. Raut, S.S. Ghugre.

*IEM, Kolkata, India.*

R. Banik.

*Government General Degree College of Pedong, Kalimpong, India.*

Sajad Ali.

*Mugberia Gangadhar Mahavidyalaya, Purba Medinipur, India.*

W. Shaikh.

*University of Calcutta, Kolkata, India.*

G. Gangopadhyay.

[1]sukalyan.chattopadhyay@saha.ac.in

## Contributions

S.C., P.D. and A.K. conceived the work. The multi-detector array was set up and kept operational by a team comprising of Soumik Bhattacharya, S.Dar, S.Basu, G.M., S.B., A.K., P.D., S.N., S.S.N., S.D., R.B., S.A., W.S.. The DAQ was set-up by R.R. and S.G.. The data analysis and interpretation were carried out by A.K., S.C. P.D. and H.P. The theoretical calculations and interpretation were performed by G.G. and S.C.. The manuscript was prepared by S.C., A.K. and P.D. in consultation with all other authors and their comments have been incorporated. All authors except G.G. took part in the experiment.

## Corresponding author

Correspondence to S.C.

Email: sukalyan.chattopadhyay@saha.ac.in